\documentclass[10pt,letterpaper,twocolumn]{article} 

\usepackage{ol2}
\usepackage[draft,implicit=false]{hyperref}
\usepackage{amsmath}

\usepackage{epstopdf}
\usepackage{hyperref}

\begin{document}

\twocolumn[ 

\title{Enhancing infrared extinction and absorption in a monolayer graphene sheet by harvesting the electric dipolar mode of split ring resonators}

\author{Yuancheng Fan,$^{1,*}$ Zeyong Wei,$^{1}$ Zhengren Zhang,$^{1}$ and Hongqiang Li$^{1,2}$}

\address{
$^1$Key Laboratory of Advanced Micro-structure Materials (MOE) and\\
School of Physics Science and Engineering, Tongji University, Shanghai 200092, China\\
$^2$e-mail: hqlee@tongji.edu.cn\\
$^*$e-mail: 429yuancheng@tongji.edu.cn, boyle429@126.com
}

\begin{abstract}Optical extinction and absorption enhancement in the infrared range of a monolayer graphene sheet by patterning split ring resonators (SRRs) is studied. It is found that the electric mode is stronger in enhancing infrared extinction and absorption compared to the magnetic mode and other higher-order modes. We improve the infrared extinction of the SRR graphene sheet by increasing the graphene area ratio in the SRR unit cell design. With the increase of the graphene area ratio, the radiation ability of the electric dipolar mode and dissipation of graphene compete for a maximum infrared absorption of about 50\%. The findings on enhancing infrared extinction and absorption of graphene sheet by harvesting the electric dipolar mode may have potential applications in terahertz and infrared detection and modulation for graphene photonics and optoelectronics.
\end{abstract}


 ] 

Graphene, a monolayer carbon atoms in hexagonal lattice with high-performance electric, thermal and mechanic properties \cite{1}, has recently attracted considerable attention for both its fundamental physics and enormous applications, e.g., in nanoelectronics. Such a realistic two-dimensional (2D) material also shows a rise in photonics and optoelectronics \cite{2,3}. As a new optoelectronic material, graphene exhibit a strong binding of surface plasmon polaritons and support its relatively long propagation \cite{4,5}. Linear dispersion of the 2D Dirac fermions also provides ultrawideband tunability through the electrostatic field, magnetic field or chemical doping \cite{2,6,7}. One remarkable feature of the graphene is that as an extremely thin film graphene is almost transparent to optical waves \cite{8}. However, an optical insulator \cite{9,10} similar to gapped graphene \cite{11,12,13} for nanoelectronics is frequently required in a myriad of applications for all-optical systems and components of much miniaturized optical circuits \cite{14,15}. Artificially constructed micro-structure, i.e., metamaterial \cite{16,17}, as a platform for enhancing light-matter interactions \cite{18,19} have been introduced for this purpose. A pioneering study reported that optical absorption enhancement can be achieved in periodically doped graphene nanodisks. In the work, a periodic graphene nanodisk is overlying on a substrate or on a dielectric film coating on metal. Excitation of the electric mode of the nanodisks together with multireflection from the assistants of the total internal reflection and metal reflection can result in complete optical absorption\cite{20}.

In this Letter, we propose to enhance infrared extinction and absorption in a monolayer graphene sheet by patterning split ring resonators (SRRs). By introducing asymmetric split ring resonators (ASRRs) into the monolayer graphene sheet, we excited both fundamental magnetic mode and electric mode, and the contributions on enhancement of infrared extinction and absorption of these two modes are comparatively studied. It is found that the electric mode performs better than the magnetic mode and higher order modes in strengthening infrared extinction and absorption. The designed periodic SRRs structure patterned sheet is shown in Fig. 1, square ring with width $a=2.5$ $\mu$m is patterned in a square lattice with lattice constant $P=3.0$ $\mu$m, $g=0.6$ $\mu$m is gap size of the SRR. Line-width $w$ and gap center with respect to center of the square ring $\delta$ are variable parameters which can significantly influence infrared extinction properties of the graphene sheet. The structured graphene sheet lies in the $xy$-plane, and throughout our study metamaterials are illuminated along the $z$-direction by $x$-polarized (the electric field $\textit{\textbf{E}}$ is along the $x$-axis) optical waves as illustrated in Fig. 1.
\begin{figure}[b]
\includegraphics[width=8.6cm]{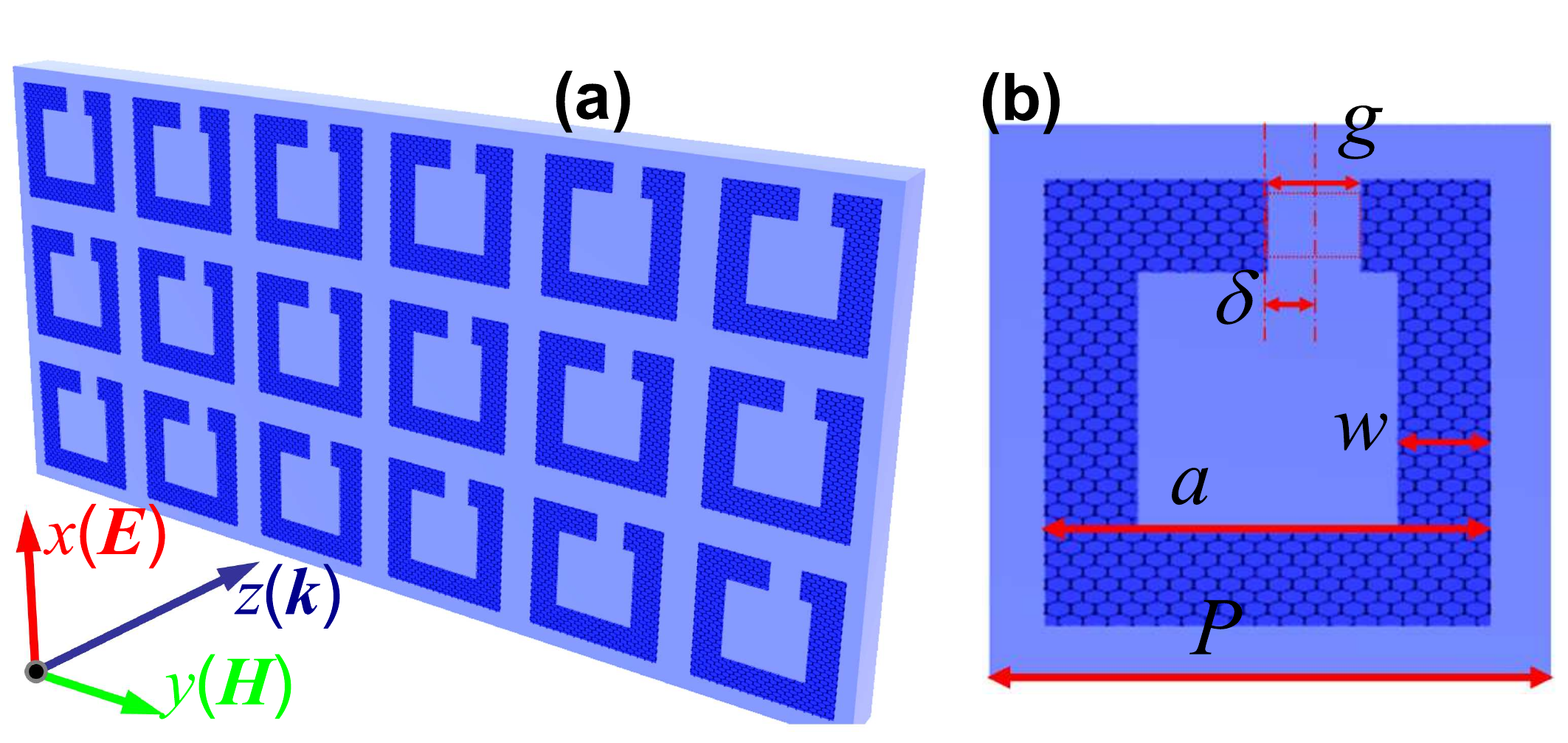}\caption{(a) 
Schematic of the SRR patterned monolayer graphene sheet and corresponding electromagnetic excitation configuration (the polarization direction is along the x axis). (b) A unit cell of the SRR patterned graphene sheet. Geometric parameters are denoted by black letters.}
\end{figure}

In calculations, the graphene sheet was approximatively treated as optical interface with complex surface conductivity, since a one-atom-thick graphene sheet is sufficiently thin compared with the concerned wavelength. Complex conductivity of the graphene is adopted from a random-phase-approximation (RPA) \cite{21,22,23}, which can be well-described by a Drude model \cite{7,24,25,26} as $\sigma=e^{2}E_{F}/\left(\pi\hbar^{2}\right)\ast i/\left(\omega+i\tau^{-1}\right)$, especially at low frequencies and for heavily doped region, where $E_{F}=0.5$ eV is the Fermi energy away from Dirac point, and $\tau=\mu E_{F}/e\upsilon_{F}^2$ is the relaxation rate with $\mu=10^{4}$ cm$^2$V$^{-1}$s$^{-1}$ and $\upsilon_{F}\approx10^{6}$m/s are the mobility and Fermi velocity, respectively. $E_{F}=\hbar \upsilon_{F}\sqrt{\pi\left|n\right|}$ can be easily controlled by electrostatic doping via tuning charge-carrier density $n$ \cite{26,27,28}. A commercial EM solver (CST Microwave Studio) was employed for all the numerical simulations.
\begin{figure}[ptb]
\includegraphics[width=8.6cm]{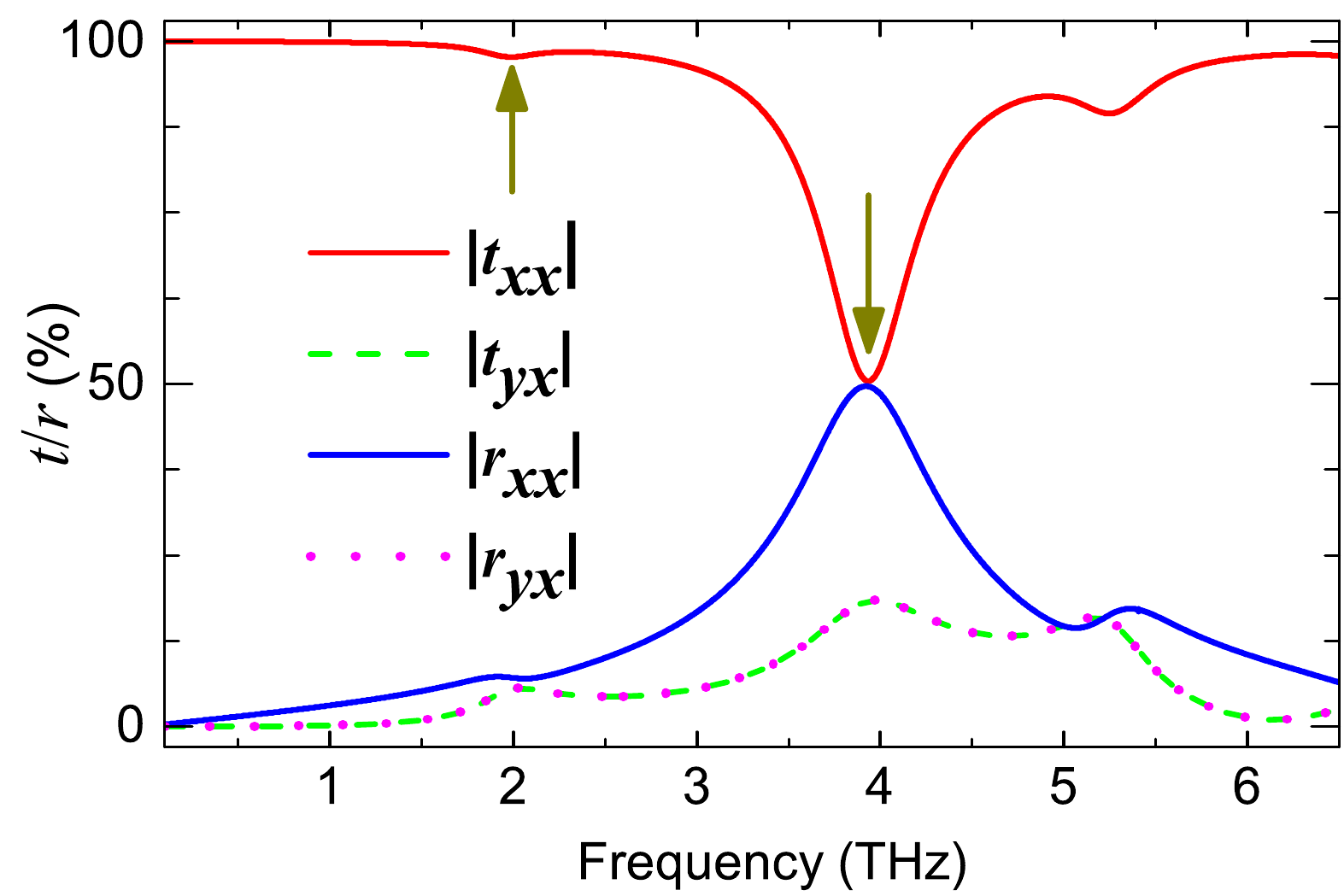}\caption{Transmission ($t$) and reflection ($r$) coefficients with respect to the $x$-polarized incident wave of ASRRs patterned graphene sheet (gap position $\delta = 0.45$ $\mu$m and line-width of graphene ASRR $w = 0.5$ $\mu$m).}
\end{figure}

We start with an asymmetric split ring resonators (ASRRs) patterned graphene sheet, with $\delta=0.45$ $\mu$m $\neq 0$ which involves asymmetry in SRRs ($w = 0.5$ $\mu$m). In the ASRR configuration we can excite the  electric mode as well as the fundamental magnetic mode at normal incidence \cite{29,30,31}. As can be seen in Fig. 2, there are two dips at low frequency (marked with arrows) on the transmission spectrum. The transmission (or reflection) coefficient of the electric-field magnitude with respect to the $x$-polarized incident wave is defined as $t_{ix} = \left|E^{Tran}_{i} / E^{Inc}_{x}\right| (i=x,y) $ (or $r_{ix} = \left|E^{Ref}_{i} / E^{Inc}_{x}\right|$), where $E^{Inc}_x$ is the electric field of the incident wave, and $E^{Trans}_{i}$ and $E^{Ref}_{i}$ are the $x$ and $y$ components of the electric field of transmitted or reflected waves. We confirmed, by numerical simulations, that these two dips corresponding to the excitation of the fundamental magnetic mode and electric mode, respectively. The reason for the higher frequency resonance at 3.96 THz is fairly obvious. This suggests that the electric resonant mode is well excited. The lower frequency mode at 1.99 THz shows a very shallow dip in the $t_{xx}$ spectrum, which implies that the magnetic mode in the graphene ASRR is weakly excited. From the resonant strength of the dips it is obvious that the electric resonant mode is stronger in enhancing light¨Cgraphene interactions and thus infra red extinction and absorption compared with the mag netic mode and other higher-order modes. We believe this is due to graphene¡¯s relatively low carrier concentra tions that lead to its weak capturing ability of infrared waves. Under this condition, the ubiquitous electric dipolar mode is easier to excite, because to excite the electric mode it is not necessary to capture the incident field and shape into a current loop. We also notice that $r_{yx} = t_{yx} \neq 0$. The polarization transformation comes from the radiation of induced asymmetric surface current of the ASRR patterned graphene sheet. The polarization transformation will also influence infrared extinction and absorption of the $x$-polarized incident waves we are concerned with in this Letter. We define the infrared extinction with respect to $x$-polarized incident wave as $E = 1- \left|t_{xx}\right|^2$ and the absorption as $A = 1 - \left|t_{xx}\right|^2 - \left|r_{xx}\right|^2 - \left|t_{yx}\right|^2 - \left|r_{yx}\right|^2$.

\begin{figure}[ptb]\includegraphics[width=8.6cm]{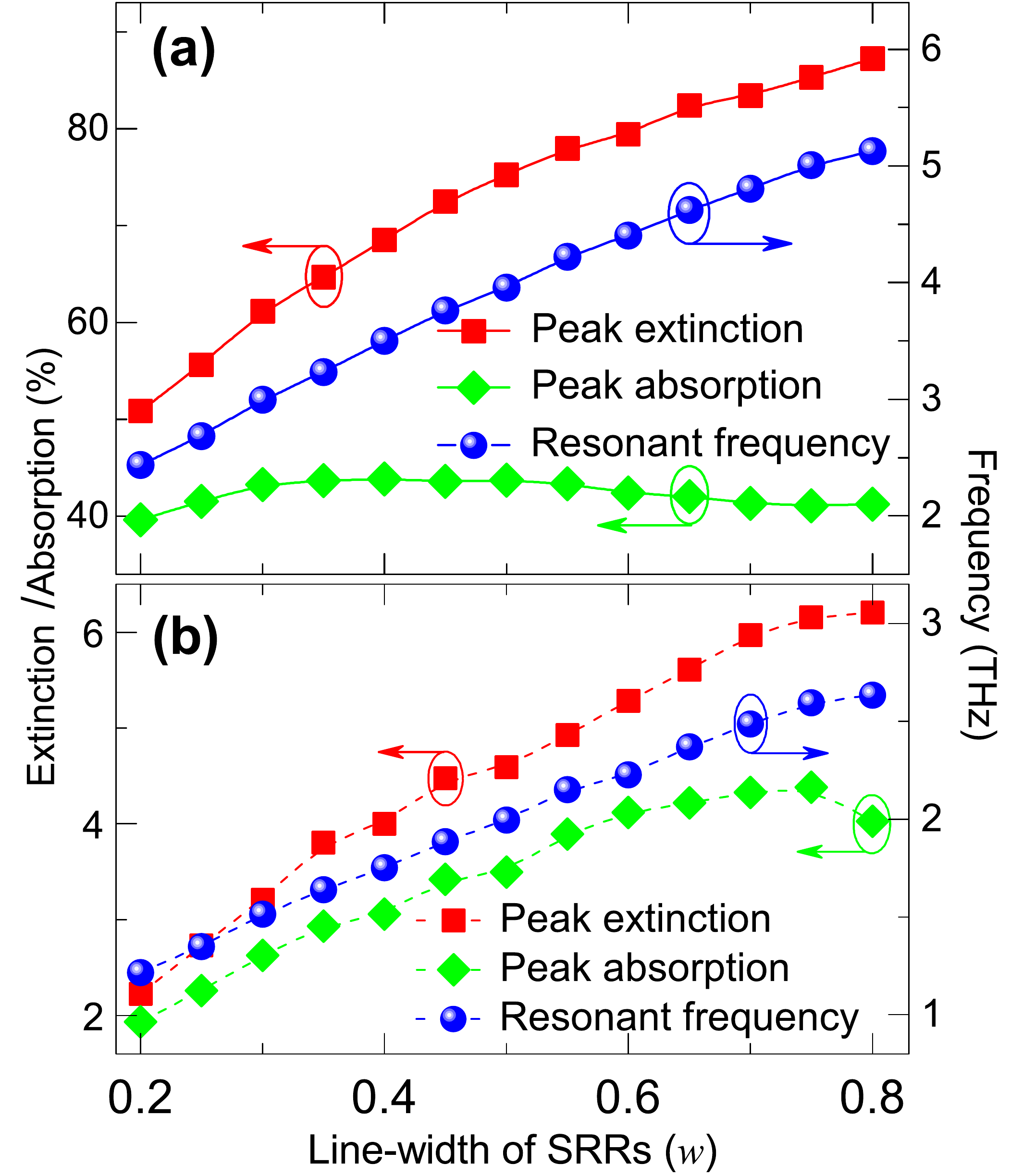}\caption{Resonant frequency (blue sphere), extinction (red square) and absorption (olive diamond) at resonant frequencies of the ASRR (with $\delta = 0.45$) patterned monolayer graphene sheet with respect to different line-width $w$ for electric mode (solid) (a) and magnetic mode (dashed) (b).}
\end{figure}

We investigate the influences of geometric parameters on the infrared extinction and absorption of the ASRR patterned graphene sheet. Figure 3 shows resonant frequencies of the electric mode and magnetic mode, and extinction and absorption at resonant frequencies for different linewidths of the ASRR. The extinction at both the electric resonant frequency and magnetic resonant frequency raises with the increase of linewidth, and the extinction at electric resonance is one-order of magnitude larger than the extinction at magnetic resonance. The infrared extinction can be efficiently enhanced at electric resonance, e.g., we can achieve an optical extinction of about 87\% at a wavelength of $58.5$ $\mu$m. It is interesting to notice the two inflexion points in the absorption curves by monotonely increasing $w$. At these points, we have maximum absorption for the ASRR patterned graphene sheet. That is 43.8\% ($w=$ 0.4$\mu$m) for the electric mode and 4.4\% ($w=$ 0.75$\mu$m) for the magnetic mode.

\begin{figure}[bt]
\includegraphics[width=8.6cm]{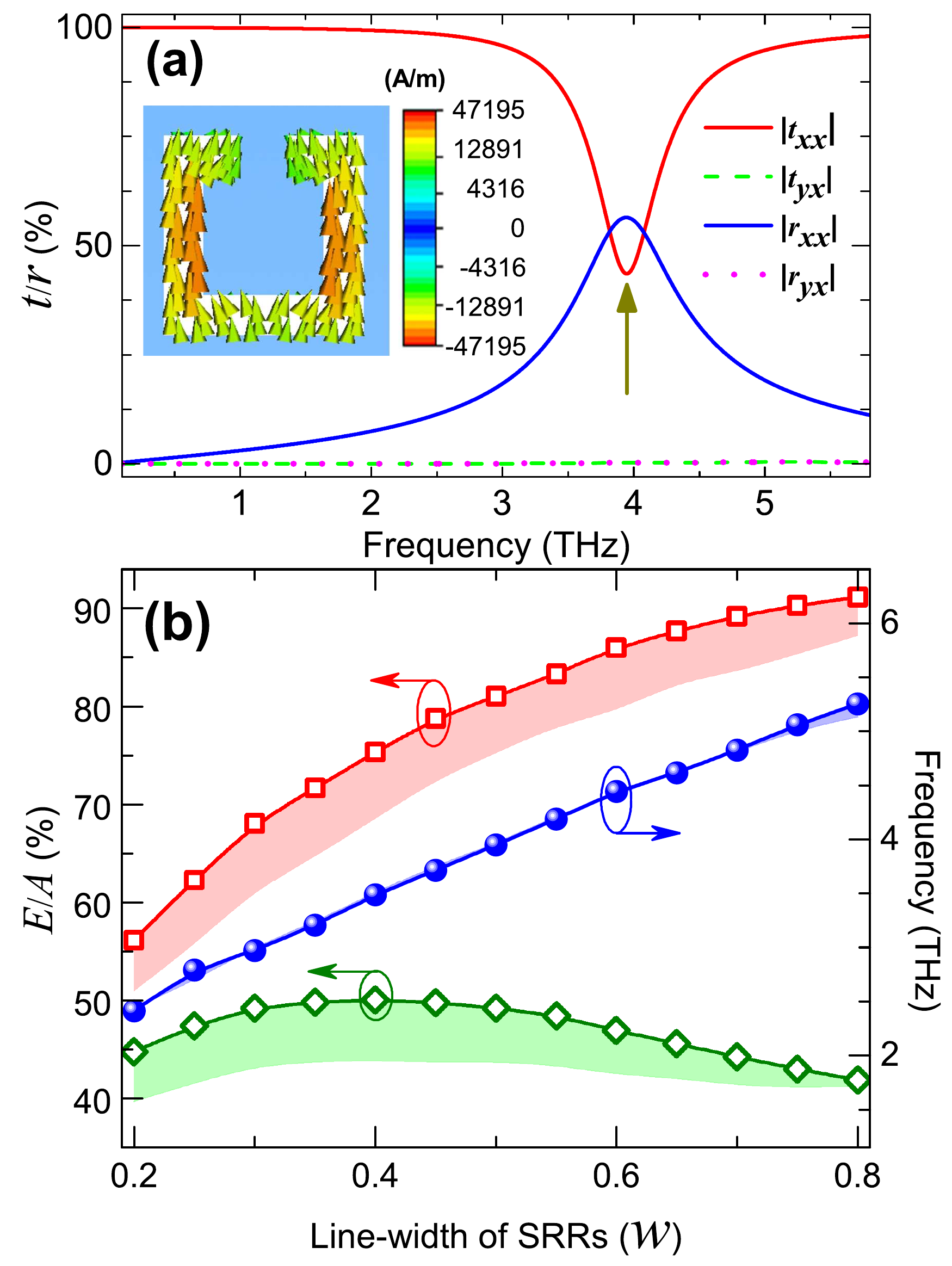}\caption{(a) Transmission and reflection coefficients of a symmetric SRR ($w=$ 0.5$\mu$m) patterned monolayer graphene sheet, the inset shows surface current distribution at the resonant frequency (marked with an arrow). (b) Extinction (red square) and absorption (olive diamond) at resonant frequencies (blue sphere) for electric mode of a symmetric SRR patterned monolayer graphene sheet with respect to different line-width $w$, the improvement of infrared extinction and absorption and corresponding resonant frequency shift by changing the symmetry of the SRR structure is indicated by the shaded area.}
\end{figure}

Then we investigated the symmetry (of the SRR structure) influence on the optical extinction and absorption. Figure 4(a) shows transmission and reflection spectra of a symmetric SRR ($w$ = 0.5$\mu$m) patterned monolayer graphene sheet. Normal incident waves can only excite the electric mode at $3.95$ THz [the inset in Fig. 4(a) shows the surface current distribution]. We find that the frequency of the electric dipole mode dips in the transmission spectra almost did not shift with changing the asymmetric parameter (with fixed $w$), and the enhancement of extinction and absorption of the symmetric SRR ($E = 81$\%, $A = 49$\%) is stronger that the ASRR ($E = 75$\%, $A = 43$\%). We compare the extinction and absorption of symmetric ($\delta = 0$$\mu$m) and asymmetric ($\delta = 0.45$$\mu$m) SRR patterned graphene sheet in Fig. 4(b). The resonant frequency, and extinction and absorption at resonant frequencies for the electric mode of a symmetric SRR patterned monolayer graphene sheet with respect to different linewidth w is shown with solid lines. The improvement in the infrared extinction and absorption and corresponding resonant frequency shift of symmetric SRR compared to ASRR by changing the symmetry of the SRR structure is indicated by the shaded area. We can see that the resonant frequencies change slightly, and symmetric SRR is better than ASRR for enhancing extinction and absorption in a graphene sheet. We have a maximum extinction of about 90\% ($w=$ 0.8$\mu$m) and a maximum absorption of 50\% ($w=$ 0.4$\mu$m) in a symmetric SRR patterned graphene sheet. Since the symmetric SRR structure is simpler [none polarization transformation, i.e., $r_{yx} = t_{yx} = 0$, in Fig. 4(a)], we try to understand the extinction and absorption enhancement processing of the symmetric SRR. First, SRRs trap optical energy that shines on graphene to their best receiving ability. Then, the trapped energy is redistributed through an absorbing channel and a radiating channel (forward and backward radiation). The receiving/radiating ability and trapped ratio with respect to incoming energy are two key points in this processing. These two aspects are both coupled to the linewidth of the SRR. On one hand, changing $w$ can significantly influence the radiating property since we know that a thin electric dipole antenna possesses better receiving/radiating ability. On the other hand, $w$ is directly linked with the graphene area ratio with respect to the unit cell which can modulate the trapped ratio as well as absorption. The electric mode and dissipation of graphene will compete for maximum infrared absorption at $w$. The incident wave ismore likely to be scattered with increased $w$, which contributes to incremental infrared extinction.

In summary, we show that infrared extinction and absorption can be enhanced in a single graphene sheet by patterning SRRs. It is found that we can significantly effect excitation of the modes of graphene SRRs by manipulating geometric symmetry of the SRRs. This shows that the electric dipole mode is stonger in enhancing infrared extinction and absorption compared to the magnetic mode and higher-order modes. We find that infrared extinction of the SRR graphene sheet can be further improved by increasing the graphene area ratio with respect to the unit cell. Meanwhile, the radiating ability of the electric dipole mode and dissipation of graphene will compete for a maximum infrared absorption of about 50\%. Tailoring scattering and absorption with the electric mode of strong resonant strength may be used in suppressing the backward scattering for ¡®cloaking¡¯ or stealth application \cite{6,32}. The electric resonant mode modulated transmission dip of a graphene sheet is beneficial for switching terahertz wave propagation \cite{33}. Our results on enhancing optical extinction and absorption of a graphene sheet with the electric resonance may have potential applications for terahertz and infrared band graphene photonics and optoelectronics.

The authors acknowledge helpful suggestions from the reviewers. This work was supported by NSFC (Grants Nos. 11174221 and 10974144), CNKBRSF (Grant No. 2011CB922001). Y. Fan acknowledges financial support from the China Scholarship Council (No. 201206260055).


\end{document}